\documentclass[twocolumn]{aastex631}

\usepackage{amsmath}
\usepackage{color}
\usepackage{url}
\usepackage{ulem}
\usepackage{multirow}
\usepackage{amsmath}
\usepackage{wasysym}

\usepackage{rotating, graphicx}

\def\apjl{ApJ}%
\def\apjs{ApJS}%
\def\ao{Appl.~Opt.}%
\def\aap{A\&A}%
\newcommand\C   {\textsc{Clumpy}}

\shorttitle{The connection between millimeter and X-ray emission in accreting massive black holes}
\shortauthors{Ricci et al.}

\begin{document}

\title{A Tight Correlation Between Millimeter and X-ray Emission in Accreting Massive Black Holes from $\lesssim100$ Milliarcsecond-resolution ALMA Observations}

\author[0000-0001-5231-2645]{Claudio Ricci}
\affiliation{Instituto de Estudios Astrof\'isicos, Facultad de Ingenier\'ia y Ciencias, Universidad Diego Portales, Av. Ej\'ercito Libertador 441, Santiago, Chile} 
\affiliation{Kavli Institute for Astronomy and Astrophysics, Peking University, Beijing 100871, China}
\author[0000-0001-9910-3234]{Chin-Shin Chang}
\affiliation{Joint ALMA Observatory, Avenida Alonso de Cordova 3107, Vitacura 7630355, Santiago, Chile}
\author[0000-0002-6808-2052]{Taiki Kawamuro}
\affiliation{RIKEN Cluster for Pioneering Research, 2-1 Hirosawa, Wako, Saitama 351-0198, Japan}
\author[0000-0003-3474-1125]{George C. Privon}
\affiliation{National Radio Astronomy Observatory, 520 Edgemont Road, Charlottesville, VA 22903, USA}
\affiliation{Department of Astronomy, University of Florida, P.O. Box 112055, Gainesville, FL 32611, USA}
\affiliation{Department of Astronomy, University of Virginia, 530 McCormick Road, Charlottesville, VA 22904, USA}
\author[0000-0002-7962-5446]{Richard Mushotzky}
\affiliation{Department of Astronomy, University of Maryland, College Park, MD 20742, USA}
\affiliation{Joint Space-Science Institute, University of Maryland, College Park, MD 20742, USA}
\author[0000-0002-3683-7297]{Benny Trakhtenbrot}
\affiliation{School of Physics and Astronomy, Tel Aviv University, Tel Aviv 69978, Israel}
\author{Ari Laor}
\affiliation{Physics Department, Technion, Haifa 32000, Israel}
\author[0000-0002-7998-9581]{Michael J. Koss}
\affiliation{Eureka Scientific, 2452 Delmer Street Suite 100, Oakland, CA 94602-3017, USA}
\affiliation{Space Science Institute, 4750 Walnut Street, Suite 205, Boulder, CO 80301, USA}
\author[0000-0001-5785-7038]{Krista L. Smith}
\affiliation{Southern Methodist University, Department of Physics, Dallas, TX 75205, USA}
\author[0009-0007-9018-1077]{Kriti K. Gupta}
\affiliation{Instituto de Estudios Astrof\'isicos, Facultad de Ingenier\'ia y Ciencias, Universidad Diego Portales, Av. Ej\'ercito Libertador 441, Santiago, Chile}
\author[0009-0002-4945-5121]{Georgios Dimopoulos}
\affiliation{Instituto de Estudios Astrof\'isicos, Facultad de Ingenier\'ia y Ciencias, Universidad Diego Portales, Av. Ej\'ercito Libertador 441, Santiago, Chile}
\author{Susanne Aalto}
\affiliation{Department of Space, Earth and Environment, Onsala Space Observatory, Chalmers University of Technology, Onsala, Sweden}
\author{Eduardo Ros}
\affiliation{Max-Planck-Institut f\"{u}r Radioastronomie, Auf dem H\"{u}gel 69, 53121, Bonn, Germany}

\correspondingauthor{Claudio Ricci}
\email{claudio.ricci@mail.udp.cl}

\begin{abstract}
Recent studies have proposed that the nuclear millimeter continuum emission observed in nearby active galactic nuclei (AGN) could be created by the same population of electrons that gives rise to the X-ray emission that is ubiquitously observed in accreting black holes. We present the results of a dedicated high spatial resolution ($\sim 60-100$\,milliarcsecond) ALMA campaign on a volume-limited ($<50$\,Mpc) sample of 26 hard X-ray ($>10$\,keV) selected radio-quiet AGN. We find an extremely high detection rate (25/26 or $94^{+3}_{-6}\%$), which shows that nuclear emission at mm-wavelengths is nearly ubiquitous in accreting SMBHs. Our high-resolution observations show a tight correlation between the nuclear (1--23\,pc) 100\,GHz and the intrinsic X-ray emission (1$\sigma$ scatter of $0.22$\,dex). The ratio between the 100\,GHz continuum and the X-ray emission does not show any correlation with column density, black hole mass, Eddington ratio or star formation rate, which suggests that the 100\,GHz emission can be used as a proxy of SMBH accretion over a very broad range of these parameters. The strong correlation between 100\,GHz and X-ray emission in radio-quiet AGN could be used to estimate the column density based on the ratio between the observed 2--10\,keV ($F^{\rm obs}_{2-10\rm\,keV}$) and 100\,GHz ($F_{100\rm\,GHz}$) fluxes. Specifically, a ratio $\log (F^{\rm obs}_{2-10\rm\,keV}/F_{100\rm\,GHz})\leq 3.5$ strongly suggests that a source is heavily obscured [$\log (N_{\rm H}/\rm cm^{-2})\gtrsim 23.8$]. Our work shows the potential of ALMA continuum observations to detect heavily obscured AGN (up to an optical depth of one at 100\,GHz, i.e. $N_{\rm H}\simeq 10^{27}\rm\,cm^{-2}$), and to identify binary SMBHs with separations $<100$\,pc, which cannot be probed by current X-ray facilities.
\end{abstract}	
               

\setcounter{footnote}{0}

\medskip

\section{Introduction}

The vast majority ($\sim$90\%) of Active Galactic Nuclei (AGN) emit only faintly in the radio, and are therefore usually referred to as radio quiet (e.g., \citealp{Wilson:1995dc}). These objects typically do not show the prominent jets observed in radio-loud AGN \citep{Begelman:1984zu,Zensus:1997cm}. Radio emission is however detected almost ubiquitously in these radio-quiet AGN, and in many cases is unresolved and comes from a very compact, sub-kpc nuclear region (e.g., \citealp{Smith:2016xw,Smith:2020ok}, \citealp{Panessa:2019wn} and references therein). Similarly, studies carried out in the millimeter (mm) regime have shown the presence of a prominent nuclear emission component in radio-quiet AGN (e.g., \citealp{Behar:2015yu,Kawamuro:2022eg}). It has been argued that this component might be associated with the same region that produces the X-ray radiation universally observed in AGN (e.g., \citealp{Laor:2008nh,Inoue:2014ri,Doi:2016fk,Panessa:2019wn,Kawamuro:2022eg}): the so-called {\it X-ray corona}. 

The electrons in the corona up-scatter optical/UV photons produced in the accretion flow into the X-ray band. The heating mechanism of the corona is still debated, but it has been widely suggested that magnetic reconnection could play an important role (e.g., \citealp{Galeev:1979vh,Di-Matteo:1997lh,Merloni:2001ey,Merloni:2001ip}). The magnetised corona is expected to generate cyclo/synchrotron radiation, observable in the radio/mm band (e.g., \citealp{Laor:2008nh,Inoue:2014ri,Panessa:2019wn}). X-ray reverberation studies (e.g., \citealp{Fabian:2009nm,Emmanoulopoulos:2011uh,De-Marco:2013sb,Uttley:2014sx,Kara:2016in,Cackett:2021my}) have shown that the X-ray corona is located at a few gravitational radii\footnote{$R_{\rm g}=GM_{\rm BH}/c^2$ is the gravitational radius for a SMBH of mass $M_{\rm BH}$.} from the supermassive black hole (SMBH). The size of the corona has been found to be relatively small ($5-10\,R_{\rm g}$) from rapid X-ray variability (e.g., \citealp{McHardy:2005ax}), X-ray eclipses (e.g., \citealp{Risaliti:2007aa}), and micro-lensing studies (e.g., \citealp{Chartas:2009tn}). Coming from a compact region close to the SMBH, the coronal mm-wave synchrotron emission is expected to be self-absorbed, and it would therefore be more easily detectable in the mm than in the radio. The size ($R$) of a self-absorbed synchrotron source decreases with the frequency following $R\propto \nu^{-7/4}$, implying that the synchrotron emission from an X-ray corona sized source would peak at $\sim 100$\,GHz.
Several studies have indeed shown that the fluxes at 100 GHz systematically exceed the extrapolation of the low-frequency steep slope power-law (e.g., \citealp{Behar:2015yu,Behar:2018mz,Doi:2016fk,Inoue:2018jk}). Moreover, some models (e.g., \citealp{Raginski:2016om,Inoue:2018jk}) suggest that the coronal emission could produce flat synchrotron emission up to $\approx 300$\,GHz. This was recently corroborated by the observational study of \citet{Kawamuro:2022eg}, which found the mm-wave nuclear emission in their AGN sample to be spectrally flat at $\sim 230$\,GHz, with spectral slopes\footnote{Considering $F_\nu \propto \nu^{-\alpha_\nu}$.} of $\alpha_\nu\sim0.5$, inconsistent with what would be expected from thermal dust ($\alpha\sim-3.5$). 

A coronal origin for the mm continuum would produce a tight correlation between the continuum emission in the X-ray and the $\sim 100-200$\,GHz bands. Studying eight radio-quiet AGN observed by CARMA, \citet{Behar:2015yu} found a correlation between the 95\,GHz and 2--10\,keV luminosities. However, expanding the sample to $26$ objects, \citet{Behar:2018mz} found a large scatter in this correlation, likely due to the heterogeneity of the sample and of the physical scales probed, as well as to the low angular resolution of the mm data ($\gtrsim 1\arcsec$). Using higher resolution ($<1\arcsec$ or $<200$\,pc) ALMA 230\,GHz observations of 98 nearby AGN, \citet{Kawamuro:2022eg} found a tighter correlation between these two bands, with a typical scatter of $\sim 0.35$\,dex. The average ratio between the $\sim$100-200\,GHz and X-ray continuum is $\sim 10^{-4}$ \citep{Behar:2015yu,Behar:2018mz,Kawamuro:2022eg}. Interestingly, this relation is consistent with what has been observed in coronally active stars \citep{Guedel:1993yq}, which are magnetically heated, similarly to what is expected for AGN coronae, further supporting the idea of a coronal origin for the 100-200\,GHz continuum emission. However, about $50$\% of the AGN in the study of \citet{Kawamuro:2022eg} showed weak resolved emission at 230\,GHz, which potentially contaminates the nuclear emission and complicates its interpretation. 

With the goal of probing smaller scales in a homogeneous way, we study here the relation between X-ray and 100\,GHz emission using the results obtained by a dedicated very high-resolution ($<100$\,mas) ALMA campaign of a volume limited ($D<50$\,Mpc) sample of hard X-ray ($> 10$\,keV) selected radio-quiet AGN. Our observations probe physical scales between 1.5 and 23\,pc. Our sample covers a large range in column density, black hole masses, X-ray luminosities, Eddington ratios and star-formation rates.  Throughout the paper we adopt standard cosmological parameters ($H_{0}=70\rm\,km\,s^{-1}\,Mpc^{-1}$, $\Omega_{\mathrm{m}}=0.3$, $\Omega_{\Lambda}=0.7$).

\begin{table*}[t!]
\begin{center}
\caption{List of {\it SWIFT}/BAT AGN used in this study (see \S\ref{sect:sample} for details on the sample selection). The table reports the sources names (1, 2), their distance (3), column density (4), fluxes (5, 6) and luminosities (7, 8) in the 100\,GHz and 14--150\,keV bands.}
\label{tab:table1}
\begin{tabular}{llcccccc}
\hline
\hline
\noalign{\smallskip}
\multicolumn{1}{c}{(1)}	& \multicolumn{1}{c}{(2)}  &  (3) &  (4)  & (5) &  (6)  & (7) &  (8)    \\
\multicolumn{1}{c}{SWIFT ID}	& \multicolumn{1}{c}{Counterpart}  &  Distance &  $\log N_{\rm H}$  & $\log F_{100\rm\,GHz}$ &  $\log F_{14-150\rm\,keV}$  & $\log L_{100\rm\,GHz}$ &  $\log L_{14-150\rm\,keV}$    \\
 &    &  [Mpc] &  [$\rm cm^{-2}$]  & [$\rm erg\,s^{-1}\,cm^{-2}$] &[$\rm erg\,s^{-1}\,cm^{-2}$]  &[$\rm erg\,s^{-1}$]  &[$\rm erg\,s^{-1}$]      \\
\noalign{\smallskip}
\hline
\noalign{\smallskip}
SWIFT\,J0251.6$-$1639   & NGC\,1125   			& 48.0	   &  24.45  &  $-15.40$ &      $-10.33$  	&       38.04 &         43.11     \\
SWIFT\,J0543.9$-$2749   & MCG$-$05$-$14$-$012   			&  41.9   &  20.00  &  $\leq -16.22$  &   $-10.75$      	&   $\leq 37.10$     &      42.57        \\
SWIFT\,J0552.2$-$0727   & NGC\,2110  			& 34.3	   &  22.94  &  $-14.25$ &      $-9.55 	$ 	&       38.89 &          43.59     \\
SWIFT\,J0601.9$-$8636   & ESO\,5$-$4  			& 28.2	   &  24.29  &  $-15.72$ &      $-10.39$  	&       37.25 &         42.59     \\
SWIFT\,J0947.6$-$3057   & MCG$-$5$-$23$-$16  	& 36.2	   &  22.18  &  $-14.38$ &      $-9.76 	$ 	&       38.81 &          43.44     \\
SWIFT\,J0959.5$-$2248   & NGC\,3081  			& 32.5	   &  23.91  &  $-15.20$ &      $-9.92 	$ 	&       37.90 &          43.18     \\
SWIFT\,J1023.5+1952   & NGC\,3227  			& 23.0	   &  20.95  &  $ -15.07$ &      $-10.03$  	&       37.73 &         42.77     \\
SWIFT\,J1031.7$-$3451   & NGC\,3281  			& 48.1	   &  23.98  &  $-14.95$ &      $-9.99 	$ 	&       38.49 &          43.45     \\
SWIFT\,J1139.0$-$3743   & NGC\,3783  			& 38.5	   &  20.49  &  $-14.67$ &       $-9.80 	$ 	&       38.58 &          43.45     \\
SWIFT\,J1212.9+0702   & NGC\,4180  			& 43.1	   &  24.28  &   $-15.82$ &      $-10.73$  	&       37.52 &         42.61     \\
SWIFT\,J1225.8+1240   & NGC\,4388  			& 18.1	   &  23.52  &   $-14.80$ &      $-9.55 	$ 	&       37.80 &          43.05     \\
SWIFT\,J1239.6$-$0519   & NGC\,4593 			& 37.2	   &  20.00  &  $-14.91$ &      $-10.12$  		&       38.31 &         43.10     \\
SWIFT\,J1304.3$-$0532   & NGC\,4941  			& 20.5	   &  23.72  &  $-15.72$ &      $-10.71	$ 	&       36.98 &         41.98     \\
SWIFT\,J1305.4$-$4928   & NGC\,4945  			& 3.5	   	   &  24.60  &  $-14.09$ &      $-9.53 	$ 	&       37.07 &          41.63     \\
SWIFT\,J1332.0$-$7754   & ESO\,21$-$4  		& 41.6	   &  23.80  &  $-15.20$ &      $-10.69	$ 	&       38.11 &         42.63     \\
SWIFT\,J1335.8$-$3416   & MCG$-$6$-$30$-$15  	& 30.4	   &  20.85  &  $-15.22$ &      $-10.27$  		&       37.82 &         42.78     \\
SWIFT\,J1432.8$-$4412   & NGC\,5643  			& 12.7	   &  24.56  &  $-14.99$ &      $-10.04$  		&       37.29 &         42.25     \\
SWIFT\,J1442.5$-$1715   & NGC\,5728  			& 37.5	   &  24.16  &  $-14.83$ &      $-9.84 	$ 	&       38.40 &          43.38     \\
SWIFT\,J1635.0$-$5804   & ESO\,137$-$34  		& 34.1	   &  24.32  &  $-15.46$ &      $-10.42$  		&       37.69 &         42.72     \\
SWIFT\,J1652.0$-$5915A   & ESO\,138$-$1  		& 39.3	   &  25.00  &  $  -14.85$ &    $-9.46 	$ 	&       38.41 &          43.81     \\
SWIFT\,J1652.0$-$5915B   & NGC\,6221  		& 11.9	   &  21.15  &    $-15.49$ &    $-10.74$  		&       36.73 &         41.48     \\
SWIFT\,J1717.1$-$6249   & NGC\,6300  			& 13.2	   &  23.31  &  $-15.11$ &       $-10.03$  	&       37.21 &         42.29     \\
SWIFT\,J1942.6$-$1024   & NGC\,6814  			& 22.8	   &  20.97  &  $-15.32$ &      $-10.18$  		&       37.48 &         42.61     \\
 SWIFT\,J2035.6$-$5013   & Fairall\,346   		& 37.7	   &  23.08  &  $-15.37$ &      $-11.00$  		&       37.86 &         42.22     \\
 SWIFT\,J2201.9$-$3152   & NGC\,7172	  		& 33.9	   &  22.91  &  $-15.05$ &      $-9.96 $ 		&       38.09 &          43.17     \\
 SWIFT\,J2235.9$-$2602   & NGC\,7314	   		& 16.8	   &  21.60  &  $-15.43$ &      $-10.37$  		&       37.10 &         42.15     \\
\noalign{\smallskip}
\hline
\noalign{\smallskip}
\end{tabular}
\end{center}
\end{table*}

\section{Sample}\label{sect:sample}

The Burst Alert Telescope (BAT) instrument on board the Swift satellite has detected over 1000 nearby AGN ($z<0.1$) in the 14--195\,keV range \citep{Baumgartner:2013uq,Oh:2018ie}. This energy band is not affected by obscuration up to column densities of $\sim 10^{24}\rm\,cm^{-2}$, which has allowed BAT to detect and identify a significant number of heavily obscured, previously unknown AGN (e.g., \citealp{Ricci:2015fk}). Swift/BAT also probes a luminosity range consistent to that of the bulk of the AGN population at higher redshifts ($z\sim 1-4$; see Fig.\,5 of \citealp{Koss:2017wu}). 
The sources of our sample are part of the BAT AGN Spectroscopic Survey\footnote{http://bass-survey.com} (BASS; \citealp{Koss:2017wu,Koss:2022yy,Ricci:2017if}), that is measuring the optical spectra and multi-wavelength properties of this minimally biased sample of nearby AGN, with the goal of creating the benchmark of SMBH accretion at low redshifts. BASS provides accurate measurements of redshifts, X-ray luminosities, column densities, black hole masses ($M_{\rm BH}$) and Eddington ratios ($\lambda_{\rm Edd}$).

Our sample was drawn from the Swift/BAT 70-month catalog \citep{Baumgartner:2013uq}. We included all the radio-quiet AGN within 50\,Mpc with a declination $<10^{\circ}$ (i.e., accesible by ALMA). Radio-loudness was estimated using the ratio between the archival 1.4\,GHz luminosity and the intrinsic 14--195\,keV X-ray luminosity ($R_{\rm X}=L_{1.4\rm GHz}/L_{14-195\rm\,keV}$), following \citeauthor{Teng:2011gj} (\citeyear{Teng:2011gj}, see also \citealp{Terashima:2003bh}). Consistently with these studies, radio-quiet AGN were defined as those with $\log R_{\rm X}\leq -4.7$. We note that most of the sources in our sample actually have $\log R_{\rm X}\leq -5.2$, well below our radio-quiet limit. For the six sources for which no archival 1.4\,GHz fluxes were available we used a similar approach, considering either the 4.85\,GHz or the 843\,MHz fluxes, to verify that the sources were radio-quiet.
Our final sample consists of 26 objects, of which eight are unobscured ($N_{\rm\,H}<10^{22}\rm\,cm^{-2}$), 10 are obscured by Compton-thin material ($10^{22}\leq N_{\rm\,H}<10^{24}\rm\,cm^{-2}$), and eight are obscured by Compton-thick gas ($N_{\rm\,H}\geq 10^{24}\rm\,cm^{-2}$). The sample is a very good representation of the intrinsic column density distribution of nearby AGN (i.e. corrected for selction biases; e.g., \citealp{Ricci:2015fk,Ricci:2017rn,Ricci:2022ay}), and it covers a broad range of 14--150\,keV luminosities [$41.5 \leq \log (L_{14-150}/\rm erg\,s^{-1})\leq 44$], black hole masses ([$6 \leq \log(M_{\rm BH}/M_{\odot})\leq 9$] and Eddington ratios ($-3 \leq \log \lambda_{\rm Edd}\leq 0$). 

\begin{figure*}
\centering
\includegraphics[width=0.49\textwidth]{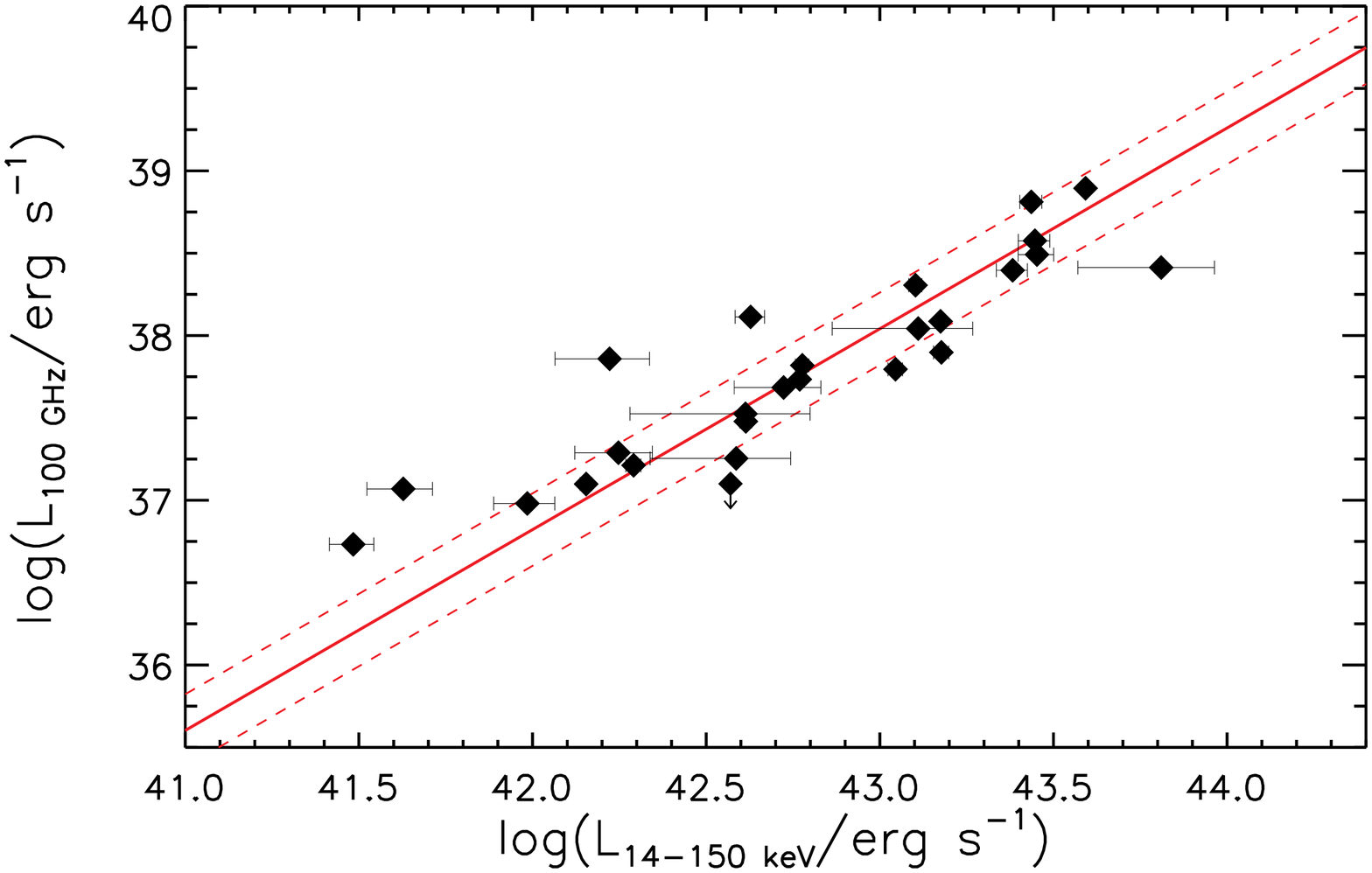} 
\includegraphics[width=0.49\textwidth]{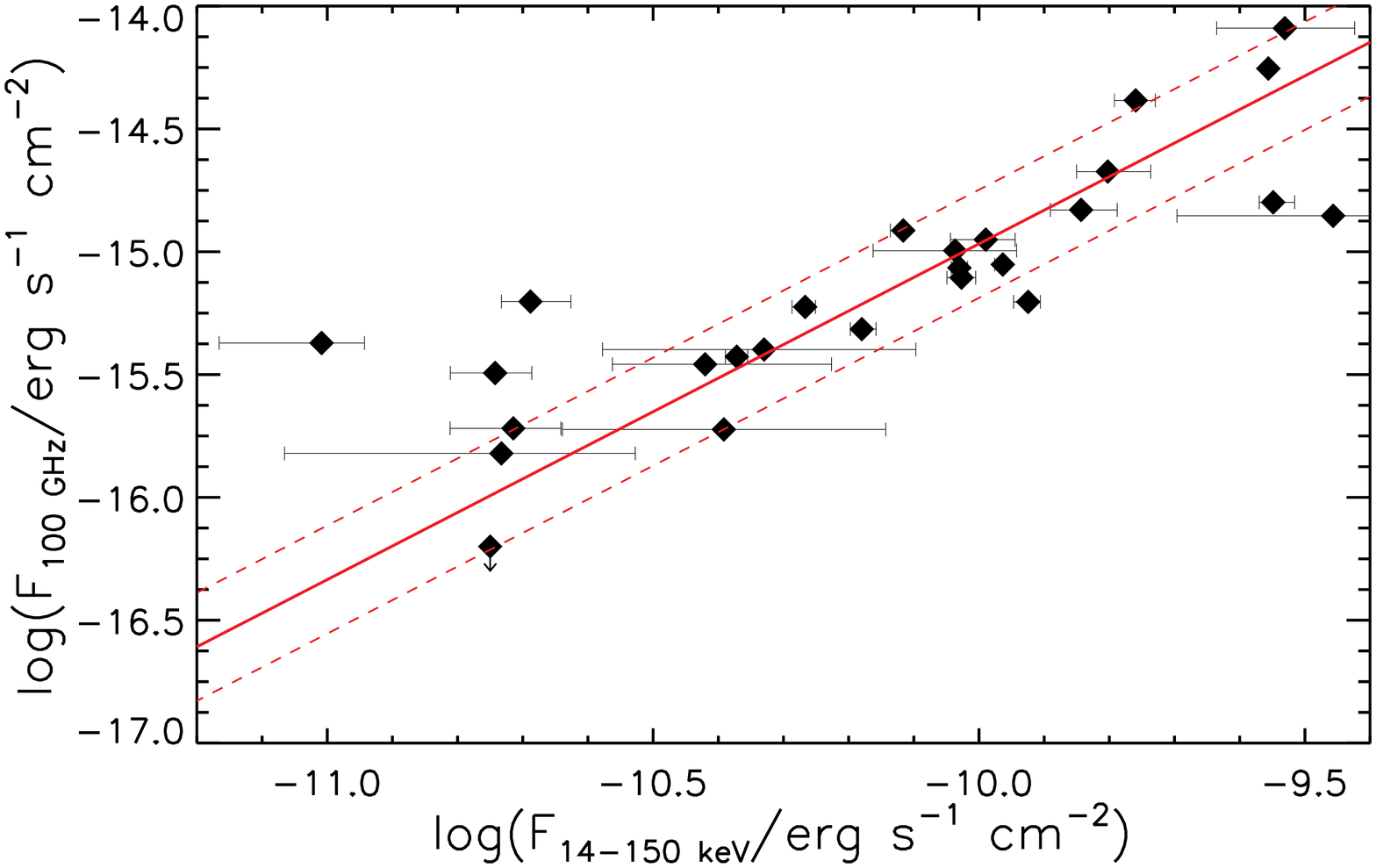} 
\par\smallskip
\caption{{\it Left panel:} 100\,GHz continuum luminosity versus the intrinsic 14--150\,keV luminosity for the sources in our sample. The solid red line represents the best fit to the data (Eq.\,\ref{eq:lum14150}), while the dashed red lines show the 1$\sigma$ scatter (0.22\,dex). {\it Right panel:} same but for the corresponding fluxes (best-fit relation given in Eq.\,\ref{eq:flux14150}). Uncertainties on the ALMA 100\,GHz fluxes and luminosities are $5\%$.}
\label{fig:Lum_lum}
\end{figure*}

We use the X-ray fluxes and column densities obtained by \cite{Ricci:2017if} through broad-band (0.3--150\,keV) X-ray spectral analysis. For the eight Compton-thick (CT) sources, we included more recent NuSTAR data to obtain a more accurate estimate of their intrinsic fluxes. We use the black hole masses reported in the second data release of BASS \citep{Koss:2022bb,Koss:2022vc,Mejia-Restrepo:2022mt}.  For the eight CT AGN, we estimated the Eddington ratios considering the intrinsic X-ray luminosity obtained by our new X-ray spectroscopic analysis (i.e., including the new NuSTAR data). This was done using the same X-ray spectral models outlined in \cite{Ricci:2017if}. For NGC\,5643 we used the results of the detailed study carried out by \cite{Annuar:2015rr}, which used a similar spectral decomposition approach. All luminosities were calculated using the distances reported in \citet{Koss:2022bb}, which include redshift-independent distance measurements for several of these nearby AGN. Bolometric luminosities ($L_{\rm Bol}$) were calculated considering a uniform 14--150\,keV bolometric correction of $\kappa_{14-150}=8.48$ ($L_{\rm Bol}=\kappa_{14-150}\times L_{14-150}$), equivalent to a 2--10\,keV bolometric correction of $\kappa_{2-10}=20$ \citep{Vasudevan:2009ng} for the median X-ray photon index of nearby AGN ($\Gamma=1.8$, \citealp{Ricci:2017if}). The Eddington ratios ($\lambda_{\rm Edd}=L_{\rm Bol}/L_{\rm Edd}$) were estimated by calculating the Eddington luminosity $L_{\rm Edd}=\frac{4\pi G M_{\rm BH} m_{\rm p}c}{\sigma_{\rm T}}$, where $G$ is the gravitational constant, $M_{\rm BH}$ is the black hole mass, $m_{\rm p}$ is the mass of the proton, $c$ is the speed of light, and $\sigma_{\rm T}$ is the Thomson cross-section. 

\section{ALMA data reduction}\label{sect:ALMAdatareduction}
The 100\,GHz fluxes were obtained by a dedicated ALMA campaign (2019.1.01230.S; PI: C. Ricci) with $\sim 60-100$\,mas resolution. We report here some basic details of our analysis. A detailed description of the ALMA analysis will be reported in a forthcoming paper (Chang et al. in prep.), which will also focus on a second ALMA campaign, that observed the same objects with lower ($0.2\arcsec$ to $0.3\arcsec$) resolution. The band-3 ALMA data was calibrated and imaged using ALMA pipeline version 2020.1.0.40 and the Common Astronomy Software Applications (CASA) version 6.1.1.15. Briggs weighting with robust parameter equals to 0.5 is used for the imaging of all the sources, with a little loss in sensitivity from robust = 2.0, which yields the possible lowest noise level, allowing us to retain almost the highest resolution. The corresponding aggregated continuum maps, and the images of the four spectral windows are created, and the primary-beam corrections are applied. For each source, the peak flux of the aggregated continuum map is used for this study. In two cases (NGC\,3281 and NGC\,4941) ALMA detected two nuclear sources (per galaxy), separated by $\sim 0.3\arcsec$. In both cases, the fluxes of the two nuclear sources were almost identical, with flux ratios of $1.07$ and $1.11$ for NGC\,4941 and NGC\,3281, respectively. For the present analysis, we used only the brighter source in each of these two systems, but we stress that this choice does not significantly change our results. Only in one case (MCG$-$05$-$14$-$012), ALMA did not detect any source at the location of the AGN, and we only report the $3\sigma$ upper limit on 100\,GHz flux and luminosity. This is likely due to the low flux of the source, as the 14--150\,keV flux of MCG$-$05$-$14$-$012 is in fact the second lowest in the sample ($17.9\times10^{-12}\rm\,erg^{-1}\,s^{-1}\,cm^{-2}$). Interestingly, this source was also not detected by ALMA at 200\,GHz \citep{Kawamuro:2022eg}. We checked whether the ALMA non-detection could be ascribed to variability, and found that a {\it NuSTAR} observation carried out $\sim3$\,weeks before the ALMA observation (ID 60160243002) points towards the same 14--150\,keV flux reported by \citeauthor{Ricci:2017if} [\citeyear{Ricci:2017if}; $\log (F_{14-150\rm\,keV}/\rm\,erg^{-1}\,s^{-1}\,cm^{-2})=-10.75$]. This seems to rule out variability as the main cause of the ALMA non detection. The 100\,GHz emission in all other objects is dominated by an unresolved nuclear component.
The sources in our sample, together with their distances, column densities, 100\,GHz and 14--150\,keV fluxes and luminosities, are listed in Table\,\ref{tab:table1}.

\bigskip
\bigskip

\section{The relation between X-ray and 100\,GHz emission}\label{sect:Xvs100GHz}

Our sensitive 100\,GHz ALMA observations detect a very high fraction (25/26 or $94^{+3}_{-6}\%$) of the radio-quiet AGN of our sample, showing that an unresolved core at mm-wavelengths is almost ubiquitous in accreting SMBHs. Interestingly, a similar detection fraction was obtained for BAT AGN at 22\,GHz by lower-resolution ($1\arcsec$) observations \citep{Smith:2016xw}.
In the left panel of Figure\,\ref{fig:Lum_lum}, we show the relation between the 100\,GHz continuum luminosity obtained through our high-resolution ALMA observations and the intrinsic 14--150\,keV luminosities from \cite{Ricci:2017if}. These X-ray luminosities were integrated over a period of 70\,months of BAT observations, and are therefore a good measure of the average AGN X-ray emission. The figure shows a very clear positive correlation between the X-ray and the 100\,GHz luminosity. The correlation is very significant, with a p-value $\lesssim 10^{-5}$. Given the very tight correlation between the 100\,GHz and the X-ray emission, the former could be used as a proxy for the AGN power. 

Fitting the logarithms of the luminosities (in $\rm erg\,s^{-1}$) with a linear relation, using a standard linear least-squares method, we obtain
\begin{equation}\label{eq:lum14150}
\log L_{100\rm\,GHz}=(-14.4\pm0.8)+(1.22\pm0.02) \log  L_{14-150\rm\,keV},
\end{equation}
which is shown as a red solid line in the left panel of Figure\,\ref{fig:Lum_lum}. The 1$\sigma$ scatter of the correlation is $\simeq 0.22$\,dex. Interestingly the slope and scatter are consistent with what was found by \cite{Kawamuro:2022eg} for the 230\,GHz emission of a subsample of radio-quiet AGN ($1.19^{+0.08}_{-0.05}$ and $0.23$\,dex, respectively).

A tight correlation is also obtained when using the 100\,GHz and 14--195\,keV fluxes (in $\rm erg\,s^{-1}\,cm^{-2}$; right panel of Figure\,\ref{fig:Lum_lum}), with the same 1$\sigma$ scatter (0.22\,dex). The best-fit relation is:
\begin{equation}\label{eq:flux14150}
\log F_{100\rm\,GHz}=(-1.3\pm0.4)+(1.37\pm0.04)\log F_{14-150\rm\,keV}.
\end{equation}

\begin{figure*}
\centering
\includegraphics[width=0.49\textwidth]{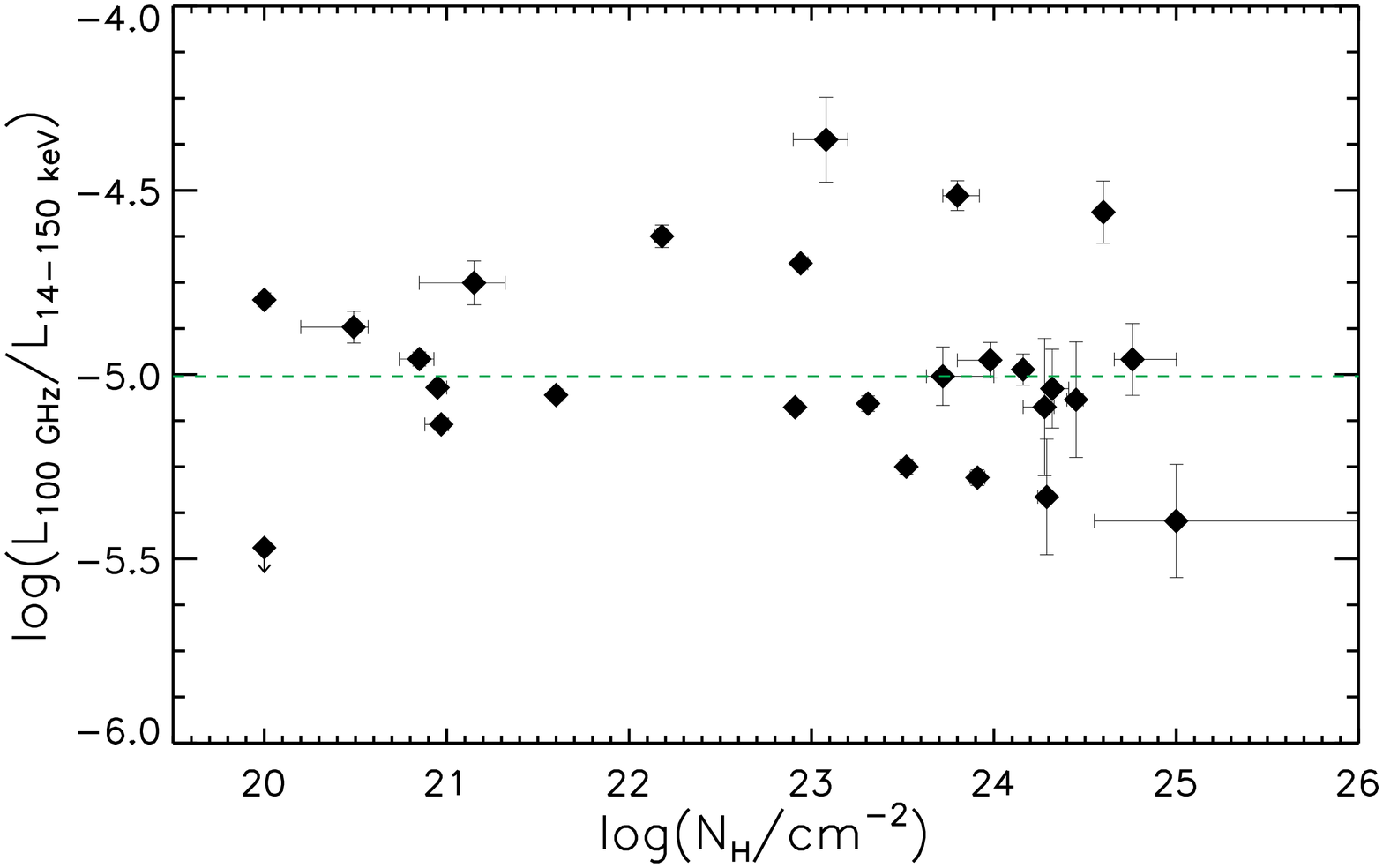} 
\includegraphics[width=0.49\textwidth]{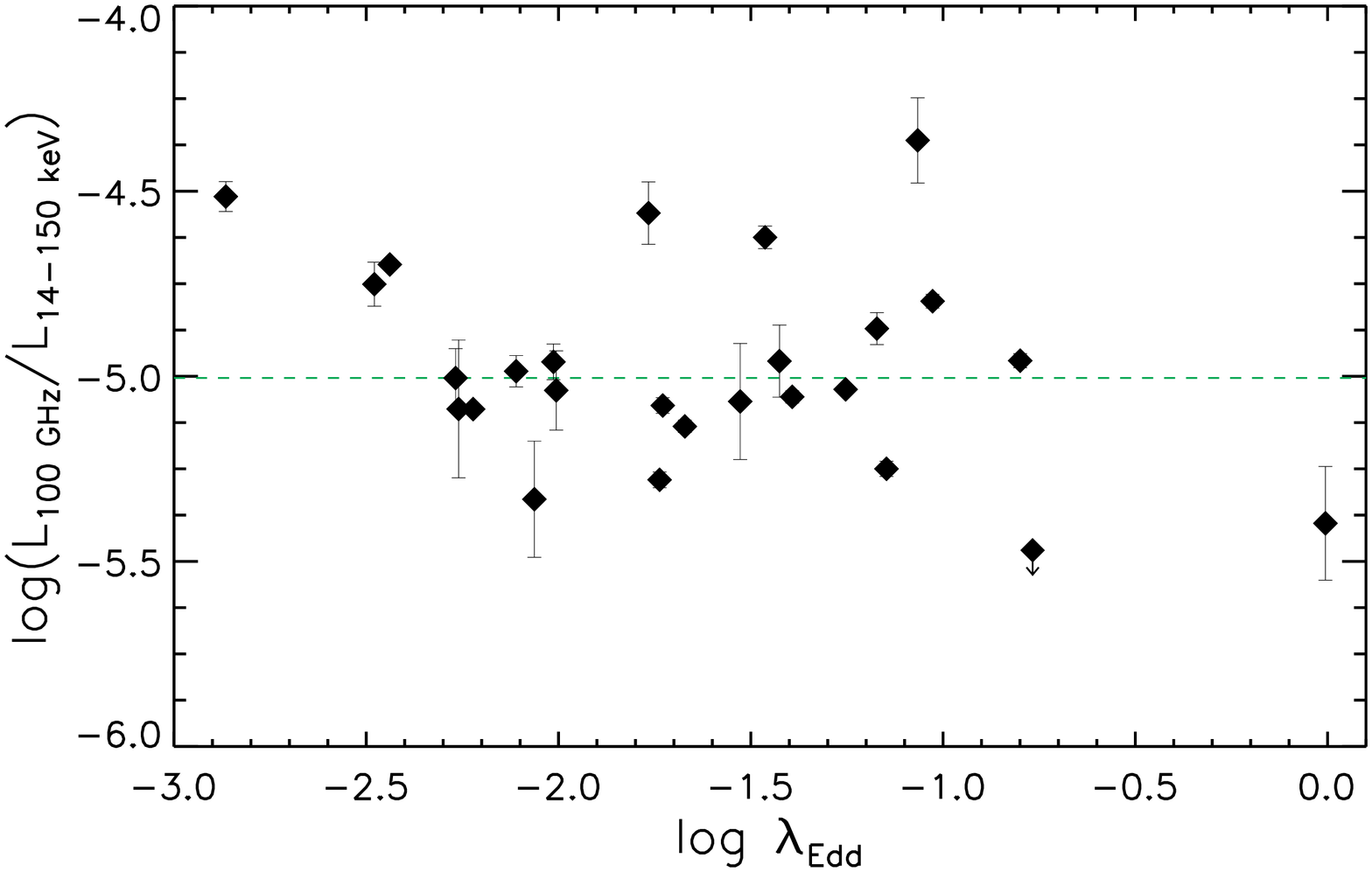} 
\includegraphics[width=0.49\textwidth]{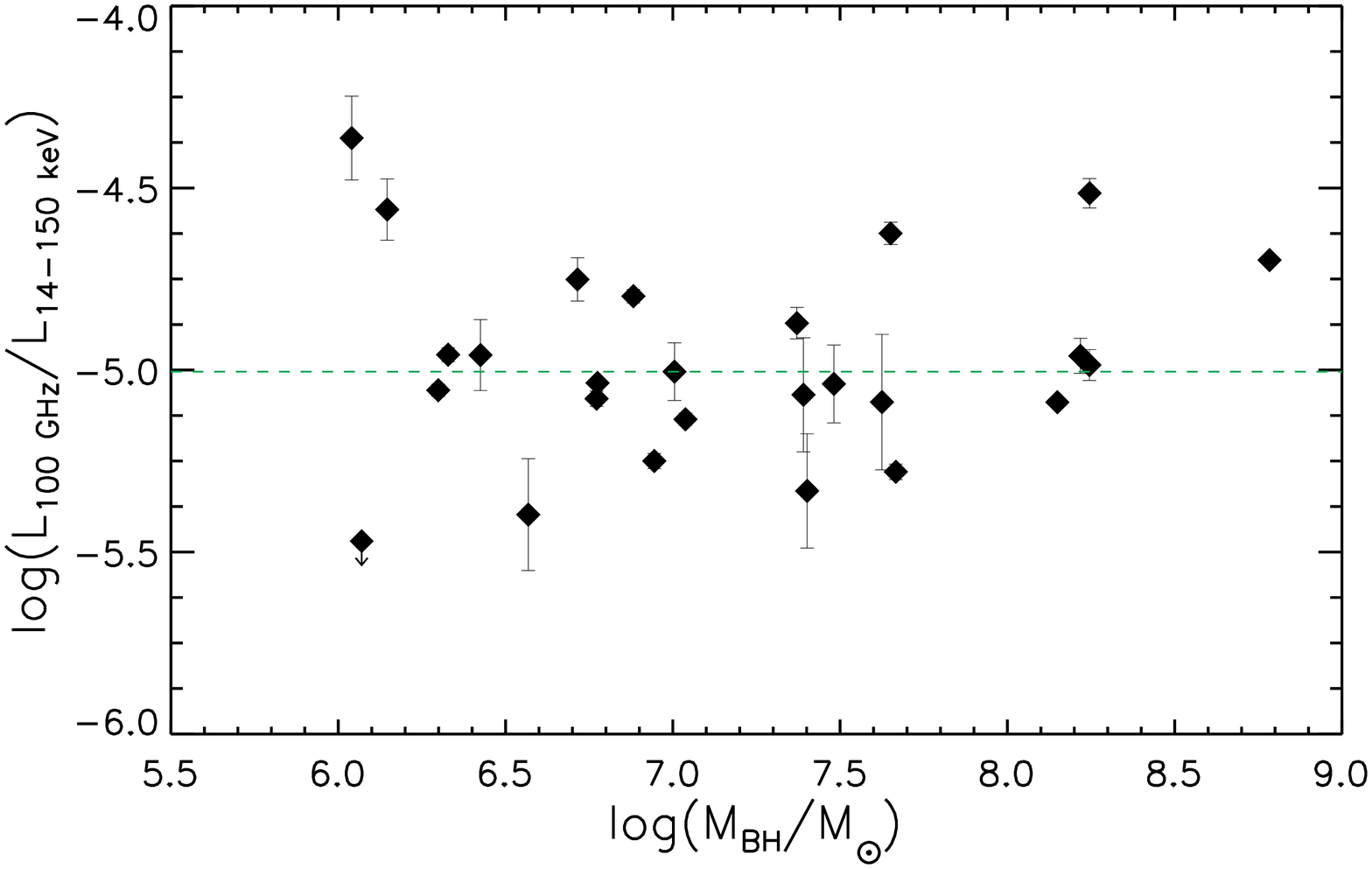} 
\includegraphics[width=0.49\textwidth]{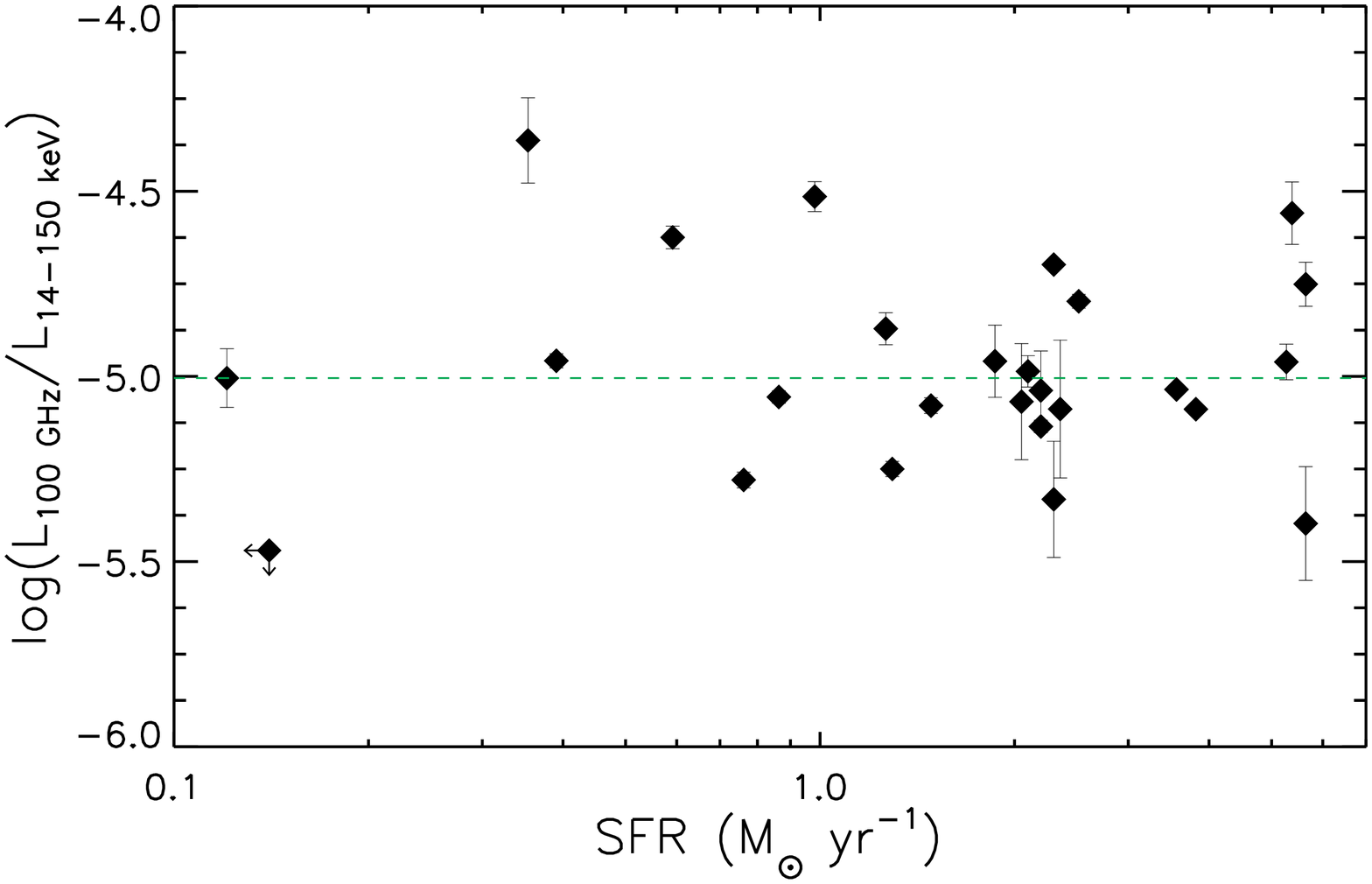} 
\par\smallskip
\caption{The ratio between 100\,GHz and intrinsic 14--150\,keV luminosities versus the column density ({\it top left panel}), Eddington ratio ({\it top right panel}), black hole mass ({\it bottom left panel}) and star formation rate ({\it bottom right panel}). The horizontal dashed green lines represent the median value of $\log (L_{100\rm\,GHz}/L_{14-150\rm\,keV})$. }
\label{fig:Ratios}
\end{figure*}

It should be noticed that, at least part of the scatter could be due to variability, since the 100\,GHz and X-ray observations are not simultaneous. Interestingly, it has been argued, by analysing the typical cross-correlation constants obtained by fitting time-averaged {\it Swift}/BAT and short {\it Swift}/XRT, {\it XMM-Newton}/EPIC and {\it Suzaku}/XIS observations, that on timescales of days to several years the X-ray variability of nonblazar AGNs is $\sim 0.2$\,dex \citep{Ricci:2017if}.

The 2--10\,keV band is commonly used in AGN surveys and studies, making it perhaps a more useful proxy of AGN coronal emission.
In order to obtain the relation between 100\,GHz and 2--10\,keV emission we first converted the 14--150\,keV fluxes and luminosities to the 2--10\,keV band assuming a pure power-law spectral model with a photon index $\Gamma=1.8$ \citep{Ricci:2017if}. This was done to consider the average 2--10\,keV emission for these objects. We then obtained a best-fit relation for the luminosities of the form:
\begin{equation}\label{eq:lum210}
\log L_{100\rm\,GHz}=(-13.9\pm0.8)+(1.22\pm0.02)\log L_{2-10\rm\,keV},
\end{equation}
while for the fluxes we obtained:
\begin{equation}\label{eq:flux210}
\log F_{100\rm\,GHz}=(0.6\pm0.4)+(1.37\pm.04)\log  F_{2-10\rm\,keV}.
\end{equation}

The typical ratio between the 100\,GHz continuum and the 14--150\,keV (2--10\,keV) emission is $\log(L_{100\rm\,GHz}/L_{14-150\rm\,keV})=-5.00\pm0.06$ [$\log(L_{100\rm\,GHz}/L_{2-10\rm\,keV})=-4.63\pm0.06$].

We further looked for possible links between the X-ray--to--mm correlation and several other key AGN quantities. Specifically, we looked at the ratio between the X-ray--to--100\,GHz continuum luminosity versus the column density (top left panel of Fig.\,\ref{fig:Ratios}), Eddington ratio (top right panel) and black hole mass (bottom left panel). We also checked how this ratio varies with star formation rate (SFR) of the host galaxy (bottom right panel of Fig.\,\ref{fig:Ratios}), inferred through IR spectral energy distribution decomposition by \citet{Ichikawa:2017zk,Ichikawa:2019pz}. In all cases we found no statistically significant trend in the X-ray--to--100\,GHz ratio. This shows that the relations reported in Eqs.\,\ref{eq:lum14150}--\ref{eq:flux210} is applicable over a wide range of AGN/SMBH and host properties. 
Moreover, this analysis provides further insight regarding the origin of the 100\,GHz emission. As discussed in \cite{Kawamuro:2022eg}, if the 100\,GHz emission was due to outflow-driven shocks, then one might expect an increase in $L_{100\rm\,GHz}/L_{14-150\rm\,keV}$ with increasing $\lambda_{\rm Edd}$, which is not observed here (top right panel of Fig.\,\ref{fig:Ratios}).

\section{Estimating column densities from the X-ray/100\,GHz relation}\label{sect:ratioVSNH}

The tight correlation between 100\,GHz continuum and the intrinsic X-ray emission could be used to infer the column density of heavily obscured objects, which are either only faintly detected in the X-rays or show a heavily obscured X-ray spectrum. 
Considering that $20-30\%$ of all AGNs are obscured by material with $\log (\rm N_{H}/cm^{-2})\gtrsim 24$ (e.g., \citealp{Burlon:2011dk,Ricci:2015fk,Torres-Alba:2021kl}), any independent approach to detect and survey them would clearly be very useful. 
Radiation at 100\,GHz can penetrate large columns of gas, even more than the hard X-ray emission, which is strongly affected by obscuration above $10^{24}\rm\,cm^{-2}$ (see Fig.\,1 of \citealp{Ricci:2015fk}). According to \citet{Hildebrand:1983ni}, the extinction at $\sim 100$\,GHz is $N_{\rm H}/\tau = 1.2 \times 10^{25} \times(\lambda/400\mu\rm m)^2$, which implies that the material becomes optically thick at 100\,GHz only at $\approx 10^{27}\rm\,cm^{-2}$, i.e. $\sim 3$ orders of magnitude above the optically-thick limit for the hard X-ray band.  Therefore, our calibration of the relation between the 100\,GHz and X-ray luminosities would allow us to use the nuclear 100\,GHz flux as a proxy of the intrinsic power of heavily obscured AGN, and the ratio between mm and X-ray observed fluxes to estimate absorbing column densities.

\begin{figure}[t!]
\centering
\includegraphics[width=0.49\textwidth]{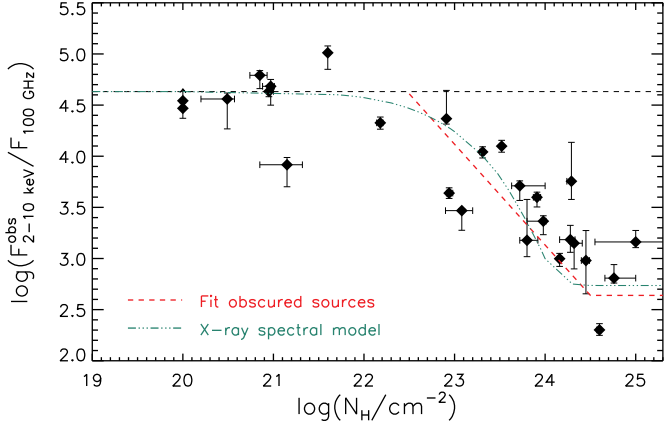} 
\par\smallskip
\caption{The ratio between the observed fluxes in the 2--10\,keV and in the 100\,GHz bands versus the column density for the objects in our sample. The horizontal black dashed line marks the median value of the ratio obtained using the intrinsic 2--10\,keV fluxes (see \S\ref{sect:Xvs100GHz}). The dashed red line represents the best fit to the data in the $\log (N_{\rm H}/\rm cm^{-2}) \simeq 22-24.5$ range (see Eq.\,\ref{eq:NHvsRatio}). The green dot-dot-dashed line show the expected decrease of the flux ratio with $N_{\rm H}$ by considering the \textsc{RXtorusD} model.}
\label{fig:RatiosvsNH}
\end{figure}

In Figure\,\ref{fig:RatiosvsNH} we show the ratio between the observed 2--10\,keV and 100\,GHz fluxes and how it varies with column density for the objects of our sample. As expected, for increasing $N_{\rm H}$ the ratio decreases, due to the increasing effect of obscuration on the observed 2--10\,keV AGN emission. The green dot-dot-dashed line in the figure shows the expected decrease of the flux ratio with column density by considering an X-ray spectral model typical of obscured AGN. This was done by using the \textsc{RefleX} \citep{Paltani:2017ev} model \textsc{RXTorusD}, the first torus X-ray spectral model that includes dusty gas \citep{Ricci:2023da}. The photon index and cutoff energy of the primary continuum in this model were set to the median values of AGN in the local universe \citep{Ricci:2017if}, i.e. $\Gamma=1.8$ and $E_{\rm C}=200$\,keV, respectively. The model includes absorbed and reprocessed radiation from a torus covering 70\% of the X-ray source, plus an unobscured Thomson scattered component with a scattered fraction of $f_{\rm scatt}=1.1\%$, which is expected to be created by material located on scales $>10-100$\,pc from the SMBH (e.g., \citealp{Bianchi:2006hj}). The latter component is responsible for the flattening of the curve for $\log (N_{\rm H}/\rm cm^{-2})\gtrsim 24.5$. It should be stressed that the behaviour of the curve for high column densities (and in particular its flattening) is significantly influenced by the choice of covering factor and scattered fraction values. A recent study focussed on nearby AGN has shown that $f_{\rm scatt}$ significantly declines for increasing column densities \citep{Gupta:2021qj}, which would lead to lower values of $F^{\rm obs}_{\rm 2-10keV}/F_{\rm 100GHz}$ for $\log (N_{\rm H}/\rm cm^{-2})\gtrsim 24.5$.

We fit our data in the $\log (N_{\rm H}/\rm cm^{-2})= 22-24.5$ interval and find that, in this range, $N_{\rm H}$ can be inferred from the ratio between the observed 2--10\,keV and the 100\,GHz fluxes ($F_{2-10\rm\,keV}^{\rm obs}/F_{100\rm\,GHz}$) using the relation:
\begin{equation}\label{eq:NHvsRatio}
\log \frac{N_{\rm H}}{\rm cm^{-2}}=(28.2\pm 0.2)+(-1.30\pm 0.05)\log  \frac{F_{2-10\rm\,keV}^{\rm obs}}{F_{100\rm\,GHz}},
\end{equation}
which is shown as a red dashed line in Fig.~\ref{fig:RatiosvsNH}. We recommend to use this relation in the range $\log (F_{2-10\rm\,keV}^{\rm obs}/F_{100\rm\,GHz})\simeq 2.7-4.5$, with $\log (F_{2-10\rm\,keV}^{\rm obs}/F_{100\rm\,GHz})\lesssim 2.7$ typically implying $\log (N_{\rm H}/\rm cm^{-2})\gtrsim 24.5$. On the other hand, $\log (F_{2-10\rm\,keV}^{\rm obs}/F_{100\rm\,GHz})\gtrsim 4.5$ implies $\log (N_{\rm H}/\rm cm^{-2})\lesssim 22.5$. In the figure for $\log (N_{\rm H}/\rm cm^{-2})\gtrsim 24.5$ we assumed a constant value of $\log(F_{2-10\rm\,keV}^{\rm obs}/F_{100\rm\,GHz})\simeq 2.7$, similarly to what was found by using the theoretical X-ray spectral model, although this value could change depending on the structure and properties of the obscuring material. The $F_{2-10\rm\,keV}^{\rm obs}/F_{100\rm\,GHz}$ ratio could be used to efficiently select obscured AGN: from Fig.~\ref{fig:RatiosvsNH} it is clear that, typically, $\log (F^{\rm obs}_{2-10\rm\,keV}/F_{100\rm\,GHz})\leq 3.5$ would strongly suggest that the AGN is heavily obscured [$\log (N_{\rm H}/\rm cm^{-2})\gtrsim 23.8$].

\section{Summary and conclusions}

With the goal of constraining the relation between mm and X-ray continuum emission in AGN, we have studied here a sample of 26 hard X-ray selected, radio-quiet AGN at distances $<50$\,Mpc with ALMA at $<100$\,milliarcsecond resolution (corresponding to $1.5-23$\,pc, see \S\ref{sect:sample} for details). The sources were selected from the Swift/BAT 70-month catalog \citep{Baumgartner:2013uq}, and have a large amount of ancillary data available \citep{Ricci:2017if,Ichikawa:2017zk,Koss:2022bb}. 
Our results are as follows:

\begin{enumerate}
\item Our sensitive 100\,GHz ALMA observations detect a very high fraction (25/26 or $94^{+3}_{-6}\%$) of the radio-quiet AGN of our sample, showing that an unresolved core at mm-wavelengths is almost ubiquitous in accreting SMBHs. 

\item Our observations indicate a very tight correlation between the 100\,GHz and the intrinsic X-ray emission (Fig.\,\ref{fig:Lum_lum}). The 1$\sigma$ scatter between the fluxes (or luminosities) is merely $0.22$\,dex (\S\ref{sect:Xvs100GHz}). Considering that the 100\,GHz and X-ray observations are not simultaneous, one would expect that the intrinsic scatter might be even smaller. The relations between the 100\,GHz and X-ray luminosities and fluxes are reported in Eqs.\,\ref{eq:lum14150}-\ref{eq:flux210}.

\item The median ratio between the 100\,GHz continuum and 14--150\,keV (2--10\,keV) emission is $\log(L_{100\rm\,GHz}/L_{14-150\rm\,keV})=-5.00\pm0.06$ [$\log(L_{100\rm\,GHz}/L_{2-10\rm\,keV})=-4.63\pm0.06$]. This ratio shows no correlation with column density, black hole mass, Eddington ratio or star formation rate (Fig.\,\ref{fig:Ratios}), which suggests that the 100\,GHz emission can be used as a proxy of the intrinsic X-ray luminosity over a broad range of these parameters (\S\ref{sect:Xvs100GHz}).

\item The tight correlation between 100\,GHz and X-ray emission could be used to infer the column density in radio-quiet AGN. The value of $N_{\rm H}$ can be inferred from the ratio between the observed 2--10\,keV and 100\,GHz fluxes using Eq.\,\ref{eq:NHvsRatio} for $\log (F_{2-10\rm\,keV}^{\rm obs}/F_{100\rm\,GHz})\simeq 2.7-4.5$ (see Fig.\,\ref{fig:RatiosvsNH}). A value of $\log (F_{2-10\rm\,keV}^{\rm obs}/F_{100\rm\,GHz})\lesssim 2.7$ typically suggests $\log (N_{\rm H}/\rm cm^{-2})\gtrsim 24.5$, while $\log (F_{2-10\rm\,keV}^{\rm obs}/F_{100\rm\,GHz})\gtrsim 4.5$ implies $\log (N_{\rm H}/\rm cm^{-2})\lesssim 22.5$. Generally $\log (F^{\rm obs}_{2-10\rm\,keV}/F_{100\rm\,GHz})\leq 3.5$ strongly suggests that the source is heavily obscured [$\log (N_{\rm H}/\rm cm^{-2})\gtrsim 23.8$; \S\ref{sect:ratioVSNH}].
\end{enumerate}

Our work shows that the nuclear 100\,GHz emission could be used as a proxy of the intrinsic (bolometric) power of accreting SMBHs. ALMA continuum observations could be very useful to detect heavily obscured AGN, even at $z\sim 1-2$, up to column densities of $\approx 10^{27}\rm\,cm^{-2}$, above which the 100\,GHz emission is also significantly attenuated. 
Moreover, these mm observations can potentially reach spatial resolutions $\sim$10--20\,times better than the best X-ray facilities (e.g., Chandra 0.5\arcsec), to identify close dual AGN in the final phase of dynamical friction (e.g., $\lesssim 250$\,pc, \citealp{Koss:2023ch}) and potentially even binary SMBHs ($\lesssim 100$\,pc).

However, it is still unclear what physical mechanisms produces the mm-continuum in AGN. It has been proposed that this emission might be produced by the X-ray corona (e.g., \citealp{Laor:2008nh,Inoue:2014ri,Behar:2018mz,Inoue:2018jk,Kawamuro:2022eg}). While the recent work of \cite{Kawamuro:2022eg} (focused on 230\,GHz emission) showed that a dust origin for the mm emission appears unlikely, one cannot rule out that it is associated with shocks produced by outflows or to free-free emission in the inner regions of the AGN. However, the lack of a correlation between $\log (L^{\rm obs}_{2-10\rm\,keV}/L_{100\rm\,GHz})$ and $\lambda_{\rm Edd}$, found both here (Fig.~\ref{fig:Ratios}) and in \citet{Kawamuro:2022eg}, appears to argue against the mm emission being associated with shocks produced by outflows, at least for the Eddington ratio regime probed here ($\lambda_{\rm Edd}\simeq 10^{-3}-10^{-0.8}$). Future studies of mm variability, correlated mm and X-ray variability, as well as higher spatial resolution studies carried out with the Global 3mm VLBI Array, will help shed light on the origin of the nuclear mm emission in AGN and its relation to the X-ray corona.

\begin{acknowledgments}
C.R. acknowledges support from the Fondecyt Regular grant 1230345 and ANID BASAL project FB210003.  B.T. acknowledges support from the European Research Council (ERC) under the European Union's Horizon 2020 research and innovation program (grant agreement 950533) and from the Israel Science Foundation (grant 1849/19). T.K. is supported by JSPS KAKENHI grant No. 23K13153 and acknowledges support by the Special Postdoctoral Researchers Program at RIKEN. SA acknowledges support from ERC Advanced grant 78941.

This paper makes use of the following ALMA data: ADS/JAO.ALMA\#2019.1.01230.S. ALMA is a partnership of ESO (representing its member states), NSF (USA) and NINS (Japan), together with NRC (Canada), MOST and ASIAA (Taiwan), and KASI (Republic of Korea), in cooperation with the Republic of Chile. The Joint ALMA Observatory is operated by ESO, AUI/NRAO and NAOJ.

\end{acknowledgments}

\facilities
ALMA, Swift, NuSTAR

\bibliography{AlmaXray}
\bibliographystyle{aasjournal}

 \end{document}